\documentclass[a4paper]{article}
\usepackage{epsfig%,showkeys
}
\usepackage{graphicx}  % standard LaTeX graphics tool
\usepackage{color}     % for color figures

\usepackage{color}
\date{\today}
 \normalsize

\newcommand{\bmat}{\left(\begin{array}}
\newcommand{\emat}{\end{array}\right)}
\newcommand{\be}{\begin{equation}}
\newcommand{\ee}{\end{equation}}
\newcommand{\bea}{\begin{eqnarray}}
\newcommand{\eea}{\end{eqnarray}}

\def\gtwid{\mathrel{\raise.3ex\hbox{$>$\kern-.75em\lower1ex\hbox{$\sim$}}}}
\def\ltwid{\mathrel{\raise.3ex\hbox{$<$\kern-.75em\lower1ex\hbox{$\sim$}}}}
\def\gev{{\rm \, Ge\kern-0.125em V}}
\def\tev{{\rm \, Te\kern-0.125em V}}

\def\m{\mu}

\def\t{\theta}

\def\la{\lambda}
\def\nn{\nonumber}
\def    \be            {\begin{equation}}
\def    \ee            {\end{equation}}
\def    \bea           {\begin{eqnarray}}
\def    \eea           {\end{eqnarray}}
\def\eps{\epsilon}
\def\a{\alpha}
\def\b{\beta}

\def\n{\nu}

\def\lam{\lambda}
\def\th{\theta}

\begin{document}
\renewcommand{\thefootnote}{\fnsymbol{footnote}}
%\rightline{IPPP/03/52} \rightline{DCPT/03/104}
\vspace{.3cm}

\title{\Large\bf Form Invariance and Symmetry in the Neutrino Mass Matrix  }

\author
{ \it \bf  E. I. Lashin$^{1,2}$\thanks{elashin@ictp.it}, S. Nasri$^{3}$ \thanks{snasri@uaeu.ac.ae}, E.
Malkawi$^{3}$ \thanks{EMalkawi@uaeu.ac.ae} and N.
Chamoun$^{4}$\thanks{nidal.chamoun@hiast.edu.sy} ,
\\ \small$^1$ Department of physics and Astronomy, College
of Science, King Saud University, Riyadh, Saudi Arabia, \\
\small$^2$ Ain Shams University, Faculty of Science, Cairo 11566,
Egypt.\\ \small$^3$Department of Physics, UAE University, P.O.Box 17551,
Al-Ain, United Arab Emirates. \\
\small$^4$  Physics Department, HIAST, P.O.Box 31983, Damascus,
Syria.  }

\maketitle
\begin{center}
\small{\bf Abstract}\\[3mm]
\end{center}
We present the general form of the unitary matrices keeping invariant the Majorana
neutrino mass matrix of specific texture suitable for explaining oscillation data.
In the case of the Tri-bimaximal pattern with two degenerate masses we give a specific
realization of the underlying $U(1)$ symmetry which can be uplifted to a symmetry in a
complete theory including charged leptons. For this, we present a model with three light SM-like
 Higgs doublets and one heavy Higgs triplet and find that one can accommodate the hierarchy of
the charged lepton masses. The lepton mass spectrum can also be  achieved in another model
extending the SM with three SM-singlet scalars transforming non trivially under the flavor symmetry.
We discuss how such a model has room for generating enough baryon asymmetry
through leptogenesis in the framework of Type-I and II seesaw mechanisms.
\begin{minipage}[h]{14.0cm}
\end{minipage}
\vskip 0.3cm \hrule \vskip 0.5cm
%%%%%%%%%%%%%%%%%%%%%%%%%%%%%%%%%%
\section{{\large \bf Introduction}}

The atmospheric, solar and reactor neutrino oscillations
\cite{SK,SNO,CHOOZ} have provided robust evidence that neutrinos
are massive and lepton flavors are mixed. Moreover, a number of
phenomenological ans$\rm\ddot{a}$tze of lepton flavor mixing with
two large rotation angles \cite{FX96,BM} have been proposed and
discussed \cite{Review}, in particular, the tri-bimaximal flavor
mixing pattern \cite{HPS} describes approximately well the
oscillation data. The starting point is usually the assumption
that there are only 3 neutrinos and that they are Majorana
fermions.  The most general neutrino mass matrix, in the flavor
basis where the charged-lepton mass matrix is diagonal, is then a
symmetric $3 \times 3$ matrix $M_\n$. Any model of neutrino mass
always ends up with a simplification of $M_\n$, thereby reducing
the number of independent parameters, and the results are fitted
in order to be consistent with the experimental data.

The approach of form invariance proposed by Ma \cite{maform},
substitutes this {\it ad hoc} procedure by the symmetry argument
that the neutrino mass matrix is invariant when expressed in the
flavor basis and another basis related to the former by a specific
unitary transformation $S$:
\begin{equation} \label{form} S^T {M_\n}
S = {M_\n}
\end{equation}
This, for certain $S$, imposes a particular form on $M_\n$ which
might be able to accommodate the data. The set of these $S$'s might form
discrete or continuous symmetry groups, depending on the mass spectrum, and
then one can impose this symmetry on the setup so to become the underlying symmetry for the
desired form of $M_\n$.

In this work, we seek the most general
symmetry $S$ satisfying the form invariance (Eq.~(\ref{form}), and we find it by examining the
invariance implication on the diagonalized neutrino mass
matrix $M^{diag}_\n$ since the analysis in the
latter case is simpler.% and one can then deduce the
%corresponding symmetry for the neutrino mass matrix in the flavor
%basis.

The method is applied to the phenomenologically successful tri-bimaximal pattern \cite{HPS}
and its realization in
tripartite model \cite{matripar}. When the three neutrino masses are distinct, the form
of $S$ is quite limited and has a well-defined $(Z_2)^3$ symmetry \cite{LESN1}. However,
in the special case when two
neutrino masses are almost degenerate, the symmetry is a priori isomorphic to the
abelian group $U(1)$ corresponding to a rotation in the degenerate mass eigenspace.
We find a realization of this approximate $U(1)$ symmetry and deduce the general form of the matrix
$S$ characterizing the tripartite model with two degenerate
masses, of which the $Z_3$ symmetry reported in \cite{matripar} is
a special case. Moreover, if a symmetry is behind the observed
pattern of ${M}_\nu$, then it must also apply to the charged
lepton mass matrix ${M}_l$. Following \cite{matripar}, we first
introduce three Higgs scalar doublets at the electroweak scale,
and one heavy Higgs triplet and find that the conditions on the
Yukawa couplings necessary to accommodate the $Z_3$ symmetry of
\cite{matripar} are sufficient to enforce the approximate $U(1)$ continuous symmetry, of which we characterize the
conserved current. We introduce later another model with only one Higgs doublet, but extending the standard model (SM)
by three SM-singlet scalars transforming non-trivially under the flavor symmetry. Like the first model, all patterns
of charged lepton masses can be accommodated, but moreover, we examine the possibility of the model to produce
enough barygenesis, via leptogenesis, in the framework of seesaw mechanisms.

The plan of the paper is as follows. We start by some basics
defining the notations in section 2. In section 3 we explain the
method for finding the form invariance symmetry, and we apply it to the tripartite model.
We treat in section 4 the case of almost two degenerate
masses. In section 5 we implement the symmetry in a set up
including the charged leptons with many Higgs doublets, and study the current associated with this
continuous symmetry. In section 6 we introduce another model with additional SM-singlet scalars and study the charged lepton
mass spectrum. In sections 7 and 8 we treat, within the model, the problems of generating the neutrino mass hierarchies and
 the lepton and baryon asymmetries in the framework of seesaw mechanisms. We end up by summarizing our  results in section
9.

\section{{\large \bf Basic notations}}

In the Standard Model (SM) of particle interactions, there are 3 lepton
families.  The charged-lepton mass matrix linking left-handed $(e,
\mu, \tau)$ to their right-handed counterparts is in general
arbitrary, but may always be diagonalized by a biunitary
transformation:
\begin{equation} \label{charged}
{M}_l = U^l_L \pmatrix{m_e & 0 & 0 \cr 0 & m_\mu & 0 \cr 0 & 0 &
m_\tau} (U^l_R)^\dagger.
\end{equation}
Similarly, the neutrino mass matrix may also be diagonalized by a
biunitary transformations if it is Dirac:
\begin{equation}
{M}^D_\nu = U^\nu_L \pmatrix{m_1 & 0 & 0 \cr 0 & m_2 & 0 \cr 0 & 0
& m_3} (U^\nu_R)^\dagger,
\end{equation}
or by one unitary transformation if it is Majorana:
\begin{equation}
{M}^M_\nu = U^\nu_L \pmatrix{m_1 & 0 & 0 \cr 0 & m_2 & 0 \cr 0 & 0
& m_3} (U^\nu_L)^T.
\end{equation}
 The
observed neutrino mixing matrix is the mismatch between $U^l_L$
and $U^\nu_L$, i.e.
\begin{eqnarray}
U_{l\nu} = (U^l_L)^\dagger U^\nu_L \simeq \pmatrix{0.83 & 0.56 &
<0.2 \cr -0.39 & 0.59 & 0.71 \cr 0.39 & -0.59 & 0.71} \simeq
\pmatrix{\sqrt{2/3} & 1/\sqrt{3} & 0 \cr -1/\sqrt{6} & 1/\sqrt{3}
& 1/\sqrt{2} \cr 1/\sqrt{6} & -1/\sqrt{3} & 1/\sqrt{2}}.
\end{eqnarray}
This approximate pattern has been dubbed tribimaximal by Harrison,
Perkins, and Scott \cite{HPS}.

If we work in the flavor basis where $M_l$ is diagonal, thus
$U_L^l = {\bf 1}$ be a unity matrix, and  assume the neutrinos
are of Majorana-type, then the flavor mixing matrix is simplified
to $V = U_L^\nu$, and so, with $M^{\mbox{diag}}_\n = \mbox{Diag}
\left( m_1, m_2, m_3\right)$, we have:
\bea M_\n &=& V \, M^{\mbox{diag}}_\n \, V^T
\label{star}
\eea

The tri-bimaximal neutrino mixing pattern can be obtained as follows. First,
we consider the product of two Euler rotation matrices:
\begin{eqnarray}
R_{12}(\theta_x) = \left ( \matrix{ c_x & s_x   & 0 \cr -s_x & c_x
& 0 \cr 0   & 0 & 1 \cr} \right ) \; &,&  R_{23}(\theta_y) = \left
( \matrix{ 1   & 0 & 0 \cr 0   & c_y & s_y \cr 0   & -s_y & c_y
\cr} \right ) \; ,
%       (5)
\end{eqnarray} (with $s_x \equiv \sin\theta_x$, $c_y \equiv \cos\theta_y$, and so
on). We then fix $\t_y$ to be equal to the `maximal mixing' angle $\theta_y = 45^\circ$, getting the mixing matrix:
\bea \label{xgeneral} V =  R_{23}\left(\theta_y=\frac{\pi}{4}\right)\;
R_{12}\left(\theta_x \right)  & = & \left ( \matrix{ c_x      & s_x & 0 \cr
-\frac{s_x}{\sqrt{2}} & \frac{c_x}{\sqrt{2}}   &
\frac{1}{\sqrt{2}} \cr \frac{s_x}{\sqrt{2}} &
-\frac{c_x}{\sqrt{2}}  & \frac{1}{\sqrt{2}} \cr} \right )
\eea
The neutrino mass matrix takes then the form
\bea
M_\n &=& \left(\begin {array}{ccc} A'_\n-B'_\n+C'_\n&(1/4)\,\sqrt {2}\tan(2\,\t_x)C'_\n&-(1/4)\,
\sqrt {2}\tan(2\,\t_x)C'_\n\\\noalign{\medskip}(1/4)\,\sqrt {2}\tan(2\,
\t_x)C'_\n&A'_\n+C'_\n&B'_\n-C'_\n\\\noalign{\medskip}-(1/4)\,\sqrt {2}\tan(2\,\t_x)C'_\n&B'_\n-
C'_\n&A'_\n+C'_\n\end {array}\right )
\label{eq1}
\eea
where \bea A'_\n &=& -(3/4)\,\cos(2\,\t_x){(m_2 - m_1)}+(1/4)\,{(m_2 + m_1)}+(1/2)\,{ m_3}, \nn \\
B'_\n &=&
-(1/4)\,{ (m_2 + m_1)}+(3/4)\,\cos(2\,\t_x){ (m_2 - m_1)}+(1/2)\,{ m_3} \nn \\C'_\n &=&\cos(2\,
\t_x){ (m_2 - m_1)}. \eea
In consequence, any `measurable' mixing angle $\t_x$ can be obtained in this way, however the experimentally
measured x-mixing angle in the tri-bimaximal pattern can be characterized as being the mixing angle which makes
the terms involving $C'_\n$ in eq.~(\ref{eq1}), proportional to the mass difference $m_2 - m_1$,
constitute a `democratic' perturbation on the form of the mass matrix when $m_1 = m_2$. This happens when
$\theta_x = \arctan (1/\sqrt{2}) \approx 35.3^\circ$ leading to:
\begin{eqnarray}
V_0 =  R_{23}\left(\theta_y = \frac{\pi}{4}\right) \; R_{12}
\left(\theta_x = \arctan\left({1\over \sqrt{2}}\right) \right)  & = &
\left (\matrix{ \frac{\sqrt{2}}{\sqrt{3}}      & \frac{1}{\sqrt{3}} & 0
\cr -\frac{1}{\sqrt{6}}    & \frac{1}{\sqrt{3}}   &
\frac{1}{\sqrt{2}} \cr \frac{1}{\sqrt{6}} & -\frac{1}{\sqrt{3}}  &
\frac{1}{\sqrt{2}} \cr} \right ) \; .
\end{eqnarray}
The vanishing of the (1,3) element in $V_0$ assures an exact
decoupling between solar ($\nu_e \rightarrow \nu_\mu$) and
atmospheric ($\nu_\mu \rightarrow \nu_\tau$) neutrino
oscillations, and the neutrino mass matrix of eq.~(\ref{eq1}) takes
the form
\begin{eqnarray}
M_\nu & = & V_0 \left ( \matrix{ m_1 & 0 & 0 \cr 0 & m_2 & 0 \cr 0
& 0 & m_3 \cr} \right ) V^{\rm T}_0
\nonumber \\
& = & \left ( \matrix{ A_\nu - B_\nu +C_\nu  & C_\nu & -C_\nu \cr
C_\nu & A_\nu + C_\nu & B_\nu - C_\nu \cr -C_\nu & B_\nu -C_\nu &
A_\nu +C_\nu \cr} \right ) \; , \label{tripartite}
\end{eqnarray}
with
\begin{eqnarray} \label{ABC}
A_\nu = \frac{m_3+m_1}{2},\;\;\;\;\;\;\;  B_\nu  =
\frac{m_3-m_1}{2} , \;\;\;\;\;\;\; C_\nu = \frac{m_2 - m_1}{3}
\;\; .
\end{eqnarray}
The form of eq.~(\ref{tripartite}) is, thus, phenomenologically desirable and the question arises
as to whether or not there is a guiding principle, say a symmetry, leading to it. One of the ways to have
$M_\n$ of a given form is to impose a form
invariance condition (Eq.~\ref{form}) for certain unitary
$S$, and our aim is to find the most general form for these
unitary matrices $S$, which can then be uplifted to symmetries
underlying the specific form of $M_\n$

\section{{\large \bf Determining the form invariance symmetry for the tri-bimaximal pattern -- Method}}
In order to find the symmetry that imposes
the form invariance property on a given mass matrix $M_\n$, we see that Eq.~(\ref{form}), using Eq.~(\ref{star}), is
equivalent to
\begin{equation} \label{formdiag} U^T {M^{diag}_\n} U = {M^{diag}_\n}
\end{equation}
where $U$ is a unitary matrix related to $S$ by
\bea S&=&V_0^* . \;U \; . V_0^T
\eea
Thus any `symmetry' $U$ for the diagonalized form can appear as a symmetry $S$ in the flavor
basis. Writing Eq. \ref{formdiag} as
\bea \left[\sqrt{M^{diag}_\n} . U . \left(\sqrt{M^{diag}_\n}\right)^{-1}\right]^T
. \left[\sqrt{M^{diag}_\n} . U . \left(\sqrt{M^{diag}_\n}\right)^{-1}\right]
&=& 1\eea where $
\sqrt{M^{diag}_\n} = \pmatrix{\sqrt{m_1} & 0 & 0 \cr 0 & \sqrt{m_2} & 0
\cr 0 & 0 & \sqrt{m_3}}$, we see that the general form of $U$ is
\bea U&=&
\left(\sqrt{M^d_\n}\right)^{-1}\; O \; \sqrt{M^d_\n}
\label{genu}
\eea
where $O$ is any $3\times 3$ orthogonal matrix. The set of matrices $U$ defined in eq.~(\ref{genu}) form
a group under matrix multiplication. However, the unitarity condition on $U$ imposes, for real matrices $O$, the following
`vanishing commutator' condition on $O$:
\bea
\label{condition}
\left[O,M^{diag}_\n\;\right]=0.
\eea
This condition does eliminate most members of the orthogonal group $O(3)$, except few
discrete subgroups such as $(Z_2)^3$ for the case of non-degenerate neutrino mass spectrum, as was shown in \cite{LESN1}.
However, in the case of degenerate spectrum there is room for few continuous subgroups to remain as we shall see now.

\section{{\large \bf Application to the tripartite model with two degenerate masses}}
Let us consider here the case of an almost degenerate mass spectrum $m_1 \simeq m_2$, where their actual difference
(the $C_\n$ part in Eq. \ref{tripartite}) can be treated as a perturbation originating from higher order operators, then
$M_\n$ is of the form:
\bea M_\n & = & \left (
\matrix{ A_\nu - B_\nu  & 0 & 0 \cr 0 & A_\nu  & B_\nu \cr 0 &
B_\nu  & A_\nu \cr} \right ) \; , \label{degetripartite}
\eea
with $A_\nu = \frac{m_3+m_1}{2}, B_\nu  = \frac{m_3-m_1}{2}$. Any rotation $V$ corresponding to
$\t_y$ being fixed at $\frac{\pi}{4}$ and $\t_x$ arbitrary
(Eq. \ref{xgeneral})
will diagonalize $M_\n$.

In \cite{matripar}, a symmetry ($Z_3 \times Z_2$) for the form
(\ref{degetripartite}), which determines it uniquely, was given:
\bea
S_B = \pmatrix {-1/2 & -\sqrt{3/8} & \sqrt{3/8} \cr
\sqrt{3/8} & 1/4 & 3/4 \cr -\sqrt{3/8} & 3/4 & 1/4}&:& S_B^3 =
{\bf 1},    \nn \\ S_2 = \pmatrix {-1 & 0 & 0 \cr 0 & 0 & 1 \cr 0 & 1
& 0} &:& S_2^2 = 1,
\label{masy}
\eea
However, from section $2$ we
 find that the general symmetry
enforcing the form (\ref{degetripartite}), which corresponds to two degenerate masses $m_1$ and $m_2$, would correspond,
provided $m_3$ is different from the common degenerate mass in accordance with the experimental data, to an orthogonal matrix $O$,
in Eq. (\ref{genu}) and satisfying the condition
(\ref{condition}), of the form  $O = \left(\begin{array}{cc} O_2 & \begin{array}{c}0\\0\end{array}  \\ \begin{array}{cc}0&0\end{array}& \pm 1 \end{array} \right)$ where $O_2$ is an isometry in the $x$-$y$ plane. To fix the ideas, we restrict our symmetry here to the connected
component of the unity, which are the rotations in the $x,y$-plane, and get \footnote{In general, the group $U$ would be generated by
the rotations and reflections: $U= \langle R_{12}(\th),I_x,I_y,I_z \rangle$}:
\bea U\left(\,\th\,\right) &=&
\pmatrix{c_\th & s_\th & 0 \cr -s_\th & c_\th & 0 \cr 0 & 0 & 1} \label{U group}
\eea
We get thus the general symmetry
\bea \label{generalsym}
S_\th = V\left(\th_x\right)\; U\left(\,\th\,\right)\;V\left(\th_x\right)^T&=&
\left ( \matrix{ c_\th
& \frac{s_\th}{\sqrt{2}} & -\frac{s_\th}{\sqrt{2}} \cr
-\frac{s_\th}{\sqrt{2}} & \frac{1}{2}(1+c_\th) &
\frac{1}{2}(1-c_\th) \cr \frac{s_\th}{\sqrt{2}} &
\frac{1}{2}(1-c_\th)  & \frac{1}{2}(1+c_\th) \cr} \right )\eea
Note that the mixing angle $\th_x$ can be taken arbitrary here since it is not determined in
the degenerate masses case.
We
can check that $S_\th$ determines uniquely the form
(\ref{degetripartite}), i.e. a necessary and sufficient condition for a matrix to be of the form
(eq. \ref{degetripartite}) is to be symmetric and invariant under the symmetry $S_\t$ for all angles $\t$:
 \bea \label{equivalence}\left[ \left(M=M^T \right) \wedge
\left( \forall \th,S_\th^TMS_\th=M\right) \right]  &\Leftrightarrow& \left[ \exists A,B: M=\left
( \matrix{ A - B  & 0 & 0 \cr 0 & A  & B \cr 0 & B & A \cr} \right
)\right],\eea We see also that the $Z_2$ and $Z_3$ of \cite{matripar} are
particular subgroups of this $U(1)$ symmetry, for $S_{-\frac{2
\pi}{3}} = S_B$ and $S_\pi = S_2$. The $S_\th$ is a
$3$-dimensional representation, albeit `reducible', of the group $U(1)$ in that
$S_{\th_1+\th_2}=S_{\th_1}S_{\th_2}$.

Had we dropped the requirement of the matrix $M$ being symmetric in the form invariance condition, then the symmetry $S_\t$
would impose the following form on $M$:
\bea \label{nonsym-equivalence}\left[
\left( \forall \th,S_\th^TMS_\th=M\right) \right]  &\Leftrightarrow& \left[ \exists A,B,C: M=\left
( \matrix{ A - B  & -C & C \cr C & A  & B \cr -C & B & A \cr} \right
)\right],\eea

\section{{\large \bf Lepton family symmetry in presence of many Higgs doublets}}
Any symmetry defined in the basis $(\nu_e,\nu_\mu,\nu_\tau)$ is
automatically applicable to $(e,\mu,\tau)$ in the complete
Lagrangian, and one should verify that the symmetry $S_\th$ can be
imposed in a complete theory including the charged leptons. We
follow the approach of \cite{matripar} and extend the standard
model of particle interactions to include three scalar doublets
$(\phi^0_i, \phi^-_i)$, playing the role of the ordinary SM Higgs field, and one very heavy triplet $(\xi^{++},
\xi^+, \xi^0)$. The leptonic Yukawa Lagrangian is given by
\begin{eqnarray}
\label{lagrangian}
{\cal L}_Y = h_{ij} [\xi^0 \nu_i \nu_j - \xi^+ (\nu_i l_j + l_i
\nu_j)/ \sqrt 2 + \xi^{++} l_i l_j] + f_{ij}^k (l_i \phi^0_j -
\nu_i \phi^-_j) l^c_k + H.c.,
\end{eqnarray}
where, under the $S_\th$ transformation,
\begin{eqnarray}
&& (\nu,l)_i \to (S_\th)_{ij} (\nu,l)_j, ~~~ l^c_k \to l^c_k, \\
&& (\phi^0,\phi^-)_i \to (S_\th)_{ij} (\phi^0,\phi^-)_j, ~~~
(\xi^{++}, \xi^+, \xi^0) \to (\xi^{++}, \xi^+, \xi^0).
\end{eqnarray}
This means
\bea
\label{h}
S_\th^T\; h\; S_\th &=& h
\\
\label{f}
S_\th^T\; f^k \; S_\th &=& f^k
\eea
Thus this Lagrangian has the global symmetry $U(1)_L\bigotimes S_{\theta}$,
where $U(1)_L$ is associate with total lepton number \footnote{We assign a zero lepton number to the doublets $\phi_i$, and a lepton
number of two units to the heavy triplet $\xi$.}
However, in order to avoid having Goldstone Bosons (majorons) in the theory, when $\xi^0$ gets a vacuum expectation value (vev) breaking
 spontaneously the $U(1)_L$ symmetry, we  add the following soft symmetry breaking term
\begin{eqnarray}
\delta {\cal L_Y} =  \frac{\mu_{ij}}{2}\phi_i^T \xi^\dagger i\tau_2 \phi_j +h.c.
\end{eqnarray}
where $\mu_{ij}$ is not proportional to the identity $\delta_{ij}$, so that the $U(1)$-symmetry $S_\theta$ symmetry is
broken explicitly as well in order not to have a corresponding `majoron' when the $\phi$'s take a vev. Assuming that the triplet mass square ($M^2_{\xi}$)
is positive, then the  minimization of the potential with respect to the filed $\xi$ gives
\begin{eqnarray}
<\xi> = \frac{-\mu_i \mu_j v_i v_j }{M^2_{\xi}}
\end{eqnarray}
which can be  naturally in the electron volt range for $\mu_{ij} \sim M_{\xi} \sim 10^{12}\,\mbox{GeV}$ \cite{ma03}.
The coexistence of two types of final states for $\xi^{++}$: $l_i^+ l_i^+$ from ${\cal L}_Y$ and
$\phi_i^+ \phi_i^+$ from $\delta {\cal L_{Y}}$, indicates a non-conservation of lepton number. However, one needs
also to impose $CP$ violation in out-of-thermal equilibrium decays to insure that the lepton asymmetry generated by
$\xi^{++}$ is not neutralized by the decays of $\xi^{--}$.

Now, since $h$ is a symmetric
matrix then the relations (Eqs.~(\ref{equivalence} and \ref{h}) lead to
\begin{equation}
h = \pmatrix {a-b & 0 & 0 \cr 0 & a & b \cr 0 & b & a},
\end{equation}
As to the equation (\ref{f}), it has a general solution for $f^k$
in the form (see Eq. \ref{nonsym-equivalence})
 \be
 \label{yukawa} f^k = \pmatrix {a_k - b_k & c_k & -c_k \cr -c_k & a_k
& b_k \cr c_k & b_k & a_k}
\ee
It is noteworthy that the solution
in eq.~(\ref{yukawa}) represents also the general solution of any
invariant matrix $f^k$ under $Z_3$ (i.e. $S^T_B\, f^k \,S_B = f^k$).
We thus conclude that the underlying symmetry of the Lagrangian
(Eq.~(\ref{lagrangian}) presented in \cite{matripar} is the $U(1)$-
symmetry $S_\th$, and that the phenomenological analysis therein
assuming $Z_3 \times Z_2$ symmetry
 does apply herein with the $U(1)$-symmetry .
In fact, when the Higgses get vevs we have the neutrino and charged lepton `gauge'
mass matrices as follows:
\bea
(M_\n)_{ij} &=& <\xi^0> h_{ij}
\label{nmass}
\eea
and
\bea
M_l&=& \pmatrix{(a_1 - b_1) v_1 + c_1 (v_2 - v_3)& (a_2 -
b_2) v_1 + c_2 (v_2 - v_3)& (a_3 - b_3) v_1 + c_3 (v_2 - v_3)\cr
-c_1 v_1 + a_1 v_2 + b_1 v_3 & -c_2 v_1 + a_2 v_2 + b_2 v_3 & -c_3
v_1 + a_3 v_2 + b_3 v_3 \cr c_1 v_1 + b_1 v_2 + a_1 v_3 & c_2 v_1
+ b_2 v_2 + a_2 v_3 & c_3 v_1 + b_3 v_2 + a_3 v_3}
\label{clmass}
\eea
where $v_i \equiv \langle \phi_i^0 \rangle$. The neutrino mass matrix is proportional to a single vev
and this translates the $U(1)$-symmetry $S_\th$ from the Yukawa couplings to the neutrino mass matrix.
One can arrange for the vevs and the Yukawa couplings such that $M_l$, after suitably rotating the
charged right-handed singlet leptons $l^c$, is the charged lepton mass matrix in the flavor space, where
($M_l\;M_l^\dagger$) is diagonal. For example, if $v_{1,2} << v_3$ then we have
\begin{eqnarray}
\label{clm}
 M_l &\approx& v_3 \left ( \matrix{ -c_1  & -c_2 & -c_3 \cr b_1 &
b_2 & b_3 \cr a_1 & a_2  & a_3  \cr} \right ),
\end{eqnarray}

As the determinant of this $M_l$ is proportional to:
$v_3^2\; {\bf a} \cdot \left( {\bf b} \times {\bf c}\right)$,
where ${\bf a}$ is the vector of components $a_i$ (similarly  for ${\bf b,c}$), we conclude that a non singular lepton
mass matrix should correspond to non-coplanar vectors (${\bf a,b,c}$). We get then
\bea
M_l\;M_l^\dagger
    &\approx&v_3^2
    \pmatrix {{\bf c}^2 & -{\bf c}. {\bf b} & -{\bf c}. {\bf a} \cr -{\bf c}. {\bf b} & {\bf b}^2 &
    {\bf a}. {\bf b} \cr -{\bf c}. {\bf a} & {\bf a}. {\bf b}
& {\bf a}^2},
\eea
 In order to show that $M_l$ can naturally represent the lepton mass matrix in the flavor space, let us just assume
the magnitudes of the three vectors coming in ratios comparable to the lepton mass ratios:
\bea \frac{|{\bf c}|}{|{\bf a}|} = \lam_e \equiv \frac{m_e}{m_\tau} \sim 3 \times 10^{-4} &,&
\frac{|{\bf b}|}{|{\bf a}|} = \lam_\mu \equiv \frac{m_\m}{m_\tau} \sim 6 \times 10^{-2},\eea
 This yields the squared mass matrix
to be written as:
\bea Q_\lam \equiv M_l M_l^\dagger &\approx& v_3^2\; |{\bf a}|^2
\left ( \matrix{
\lam_e^2 & -\lam_e \lam_\mu \cos\psi & -\lam_e \cos \phi \cr
-\lam_e \lam_\mu \cos\psi & \lam_\mu^2 & \lam_\mu \cos \theta \cr
-\lam_e \cos \phi & \lam_\mu \cos \theta & 1 \cr} \right ),
\label{Q}
\eea
where $\theta$, $\phi$ and $\psi$ are the angles between the pairs of vectors ${(\bf a,b)},(\bf a,c)$ and $(\bf b,c)$ respectively.
The diagonalization of $M_l M_l^\dagger$ by
means of an infinitesimal rotation amounts to seeking an antisymmetric matrix
\be
I_\eps = \left ( \matrix{
0 & \eps_1 & \eps_2 \cr -\eps_1 & 0 & \eps_3 \cr -\eps_2 & -\eps_3 & 0 \cr} \right ),
\ee
with small parameters $\eps'$s,
satisfying: \bea \left( Q_\lam + \left[Q_\lam,I_\eps \right] \right)_{ij}&=0,& i\neq j.
\eea
If we solve this equation analytically to express the $\eps$'s in terms of ($\lam_{e,\mu},\cos (\psi,\phi,\theta)$), we find, apart from
``fine tuned'' situations corresponding to coplanar vectors ${\bf a,b,c}$, that we get:
$\eps_3 \sim \lam_\mu, \eps_2 \sim -\lam_e$ and
$\eps_1 \sim -\lam_e/\lam_\mu$, which points to a consistent solution diagonalizing  $Q_\la$ close to the identity
matrix
given by $U^l_L = e^{I_\eps} \approx I + I_\eps$. For  the above numerical values
 and a common value $\pi/3$ for the angles, we get:
$m_e^2:m_\mu^2:m_\tau^2=6\times 10^{-8}:3\times 10^{-3}:1$, with the `exact' unitary diagonalizing matrix
given by:
\be
U^l_L \sim
\left ( \matrix{
1 & -1.6 \times 10^{-3} & -10^{-4} \cr
1.6 \times 10^{-3} & 1 & 3 \times 10^{-2} \cr
 10^{-4} & -3 \times 10^{-2} & 1 \cr} \right ).
 \ee
The deviations due to the rotations are generally small, but could interpret measuring a nonzero small
value of $U_{e3}$ which is restricted by the reactor data \cite{boehm} to be less than $0.16$ in magnitude.

As we said above, the phenomenological features of \cite{matripar}
assuming $Z_3 \times Z_2$ symmetry, in particular the leptonic
flavor changing decays through $\phi$ exchange, can be repeated
here with the underlying $U(1)$-symmetry. However, in contrast to
discrete symmetries, the existence of a continuous symmetry leads
to a conserved current, which we investigate now.

For illustration, let us restrict the discussion to the neutrino part. The invariance,  under $S_\th$,
of the `current-relevant' part of the Lagrangian depending on the field derivative:
\bea K_\n &=& i \bar{\nu}_k \gamma^\mu \partial_\mu \nu_k \eea
leads to the current:
\bea J^\mu_\n \equiv -i\,\frac{\partial{K_\n}}{\partial(\partial_\mu \n_j)} T_{jk} \n_k =  T_{jk} \bar{\n}_j \gamma^\mu \n_k
\label{current}
\eea
where $T_{jk}$ is the generator of the $U(1)$-symmetry
$S_\theta$: \bea T
    &=&
    \pmatrix {0 & {i\over \sqrt{2}} & {-i\over \sqrt{2}} \cr
    {-i\over \sqrt{2}} & 0 & 0 \cr
     {i\over \sqrt{2}} & 0 & 0}.
     \eea
In fact, $S(\t)$ (eq. \ref{generalsym}),
as a three-dimensional representation of the commutative $U(1)$ group, should be reduced to three
one-dimensional irreducible representations obtained by diagonalizing the matrix $S(\t)$ to get:
\bea \label{eigenfield} S(\t) &=& L  \pmatrix{1 & 0 & 0 \cr 0 & e^{-i\t} & 0 \cr 0 & 0
& e^{i\t}} L^\dagger \\ L &=& \pmatrix{0 & \frac{-i}{\sqrt{2}} & \frac{i}{\sqrt{2}} \cr \frac{1}{\sqrt{2}} & \frac{-1}{2} & \frac{-1}{2}
\cr \frac{1}{\sqrt{2}} & \frac{1}{2}
& \frac{1}{2}} = \pmatrix{V_0 & V_- & V_+}\eea
The `neutrino' eigenvectors $V_0,V_-,V_+$ (forming the columns of the matrix $L$) represent the neutrino fields with definite $S$-charges, respectively equal to $0,-1,1$. Writing the neutrino `gauge' fields $\bf\n^g = (\n_e,\n_\mu,\n_\tau)$
in terms of these definite $S$-charge fields, one can see that the `neutrino' current (Eq. \ref{current}) expresses explicitly the conservation of the $S$-charge, in that we have:
\bea
\label{neutcurrent}
J^\m_\n &=& \left( 0\, \bar{V}_0\,\gamma^\mu\, V_0 - 1\, \bar{V}_-\,\gamma^\mu\, V_-
+ 1\, \bar{V}_+\, \gamma^\mu \,V_+\right)
\eea
We have here a conserved current associated with a global continuous symmetry with
no gauge fields coupled to it. This is similar to the case of $U(1)$ baryon number conservation in the SM.

Using the tri-bimaximal matrix $U_{l\nu}$ to move from the neutrino
`gauge' states $\n^{\bf g}_{e,\mu,\tau}$ to neutrino `mass' states $\n^{\bf m}_{1,2,3}$, we can express the definite $S$-charge neutrino fields
${\bf V}=(V_0,V_-,V_+)$ in terms of the mass eigenstates: ${\bf V}=L^T.\n^{\bf g} = L^T.V_0.\n^{\bf m} $, which gives:
\bea
V_0 & = & \nu_3, \nn\\
V_- & =& \left(-{i\over \sqrt{3}} + {1\over \sqrt{6}}\right) \, \n_1 +
\left(-{i\over \sqrt{6}} - {1\over 3}\right) \, \n_2, \nn\\
V_+ & =& \left({i\over \sqrt{3}} + {1\over \sqrt{6}}\right) \, \n_1 +
\left({i\over \sqrt{6}} - {1\over 3}\right) \, \n_2.
\label{chnum}
\eea
One can see directly that the particular combination of mass eigen-states in (Eq.~\ref{chnum}) never
mix under free time evolution provided $\nu_1$ and $\nu_2$ have degenerate mass. This degeneracy has already
been shown to be a consequence of $U(1)$-symmetry $S_\th$. The same conclusion still holds if one think of the
underlying symmetry, in the degenerate two masses case, as  $Z_3 \times Z_2$ (see Eq.~(\ref{masy}), due to the compatibility of
both $S(\th)$ and $Z_2$ in that they commute and  have common eigen-states.

\section{{\large \bf Lepton family symmetry in presence of many heavy singlet scalars}}
The many Higss doublets in the previous model were introduced to accommodate the charged lepton
mass spectrum, but at the cost of inducing dangerous flavor changing neutral
currents \cite{BjorWein77}, which are difficult to be controlled. To remedy this situation, we introduce
in the present model three heavy SM-singlet scalars transforming  non trivially under the flavor symmetry, and keep the SM-Higgs $\Phi$
intact.
However, we enlarge the flavor symmetry so as to include an inversion in the flavor space, which means that the underlying
flavor symmetry, call it $S^I$, assumes now the form
\be
S^I = S \times  \langle I\rangle  \cong U(1) \times Z_2
\label{SI}
\ee
 with $S$  given in Eq.~(\ref{generalsym}), and $I = \mbox{Diag}\left(-1,-1,-1\right)$\footnote{More precisely, the group
 $U$ in Eq. \ref{U group} is now a direct product of two commuting groups: $U \equiv \langle R_{12}(\th),I\rangle \cong SO(2) \times Z_2$}.

We assume the SM Higgs $\Phi$ and the charged right-handed leptons $l^c_j$  to be
singlets under the $S^I$ symmetry, whereas the lepton
 left-doublets transform component-wise faithfully:
\begin{equation}
 L_i
\rightarrow S^I_{ij}\;L_j,
\end{equation}
with $i,j =1,2,3$.
The normal SM mass term for charged lepton,
\bea
\label{L1}
{\cal{L}}_1 &=& Y_{ij} \overline{L}_i \Phi  l^c_j,
\eea
should vanish now since the invariance under $S^I$ restricts
the Yukawa-couplings to satisfy the matrix equation:
\bea
\label{mateq}
\left(S^{I}\right)^T\cdot Y&=& Y.
\eea
which can not be met for an $S^I$-matrix with determinant equal to $-1$. It is noteworthy that had we chosen not to enlarge the flavor symmetry
then this mass term would have been allowed.

In order to generate lepton masses
we introduce three SM-singlet scalar fields, $\Delta_k$, one for each family (the
indices $k=1,2,3$ refer also to the flavors $e$, $\mu$ and $\tau$ respectively), and they
are coupled to the corresponding lepton left
doublet $L_k = \pmatrix{\nu_k \cr l_k}$ via the dimension 5 operator:
\begin{eqnarray}
{\cal{L}}_2 &=& \frac{f_{ikr}}{\Lambda} \overline{L}_i \Phi \Delta_k  l^c_r ,
\end{eqnarray}
where $\Lambda$ is a heavy mass scale. As said earlier, this {\it{ad hoc}} assumption of the coupling of charged
leptons with the additional Higgs fields via higher operators, and not through SM-like Yukawa terms, is suitable to
reduce the effects of flavor changing neutral currents. We assume the new scalars  $\Delta_k$ and the lepton left doublets
transform similarly under $S^I$, i.e.:
\begin{equation}
 \Delta_i \rightarrow S^I_{i j}\Delta_{j}.
\end{equation}
Invariance of the Lagrangian under the symmetry implies
\begin{equation}
S^I_{i\alpha}\,S^I_{k\beta}\, f_{ikr} = f_{\alpha \beta r}.
\end{equation}
which is written, in matrix form, as
\bea \label{f_r}
\left(S^I\right)^T \, f_r \, S^I =f_r , \eea
where $f_r$, for fixed $r$, is the matrix whose ($i,j$) entry is $f_{i j r}$. Noting that $I$ enters Eq. \ref{f_r}
quadratically, and thus cancels out, then Eq. \ref{nonsym-equivalence} imposes the form:
\begin{eqnarray}
\label{cly}
f_r=\left (
\matrix{ A_r - B_r   & C_r & -C_r \cr
-C_r & A_r & B_r \cr
C_r & B_r & A_r
\cr} \right ) .
\end{eqnarray}
When the fields $\Delta_k$ and $\phi^\circ$ take the vacuum expectation values (vevs)
$<\Delta_k >=\delta_k$ and $<\phi^\circ>=v$, the charged lepton mass
 matrix originating from ${\cal{L}}_2$ becomes:
\begin{equation}
\left( {\cal{M}}_l \right)_{ir} = \frac{v f_{ikr}}{\Lambda}\delta_k.
\end{equation}
  As we are concentrating on the neutrino sector without stating explicitly the $\Delta_k$ potential and since
the $S^I$ symmetry is broken by ``soft'' terms in the Higgs sector, we may assume a
$\Delta_3$-dominated pattern:
 $\delta_1,\delta_2 \ll \delta_3$, so to get the charged lepton mass matrix
 \begin{eqnarray}
 \label{clmdel}
 M_l &\approx&
 \frac{v\delta_3}{\Lambda}
 \left (
 \matrix{ -C_1  & -C_2 & -C_3 \cr
 B_1 & B_2  & B_3  \cr
 A_1  & A_2  & A_3  \cr} \right ).
\end{eqnarray}
The determinant of $M_l$ is proportional to the mixed product:
$\left(\frac{v\delta_3}{\Lambda}\right)^3 {\bf A} \cdot \left( {\bf B} \times {\bf C}\right)$,
where ${\bf A}$ is the vector of components $A_i$ (similarly  for ${\bf B,C}$), which means that
these three vectors
should not be coplanar in order to have a nonsingular lepton mass matrix. We get then
\begin{eqnarray}
\label{clmsquare}
M_l\; M_l^\dagger &\approx&
\frac{v^2\delta_3^2}{\Lambda^2}{
\left (
\matrix{ {\bf C . C} & {-\bf C . B} &  {-\bf C . A} \cr
{-\bf B . C} & {\bf B . B} &  {\bf B . A} \cr
{-\bf A . C} & {\bf A . B} & {\bf A . A} \cr} \right )},
\end{eqnarray}
In a similar way to the analysis in the previous many Higgs doublet model, we see that assuming
the magnitudes of the three vectors to come in ratios comparable to the lepton mass ratios:
\bea
{\bf C}^2:{\bf B}^2:{\bf A}^2 &\sim& m_e^2 : m_\mu^2 : m_\tau^2
\label{ratiosABC},
\eea
would imply
that the mixing $U^l_L$, making $U^l_L M_l M_l^\dagger {U^l_L}^\dagger$ diagonal, will be
naturally very close to the identity matrix with off-diagonal terms of order
($m_e/m_\mu \sim 5 \times 10^{-3}, m_e/m_\tau \sim 3 \times 10^{-4}, m_\mu/m_\tau \sim 6 \times 10^{-2}$). This
would mean again that our basis is the flavor basis to a very good approximation and that the hierarchical charged
lepton masses can be obtained from a hierarchy on the a priori
arbitrary Yukawa couplings (${\bf C}^2 \ll {\bf B}^2 \ll {\bf A}^2$).

\section{The neutrino mass matrix and type-I seesaw scenario }
In this scenario the effective light left neutrino mass matrix is generated through seesaw
mechanism as,
\be
M_\nu = - M_\nu^D\; M_R^{-1}\;  \left(M_\nu^D\right)^T,
\label{seesaw}
\ee
where $M_R$ is the heavy Majorana right handed neutrinos mass matrix, whereas the Dirac neutrino mass matrix comes from the Yukawa term:
\begin{eqnarray}
g_{ij} \overline{L}_i
\tilde{\Phi} \nu_{Rj},
\end{eqnarray}
with $\tilde{\Phi} = i \tau_2 \Phi^*$. As to the right neutrino, we will
assume that it transforms faithfully as
\begin{eqnarray}
\label{rtransform}
\nu_{Rj} \rightarrow S^I_{j\gamma} \nu_{R\gamma},
\end{eqnarray}
since, as we shall see,
this assumption will put constraints on the right Majorana mass
matrix. The invariance of the Lagrangian under $S^I$ implies in matrix form:
\begin{eqnarray}
\left(S^I\right)^T\; g\; S^I=g .
\end{eqnarray}
Again, noting that when $I$ enters here it does so quadratically, Eq. \ref{nonsym-equivalence} forces the form:
\begin{eqnarray}
\label{nDm}
M_\nu^D &=& v
\left (
\matrix{ A_D - B_D   & C_D & -C_D \cr
- C_D & A_D  & B_D  \cr
C_D & B_D & A_D  \cr} \right ).
\end{eqnarray}

As to the right-handed Majorana mass matrix, it originates from the term:
\begin{eqnarray}
\frac{1}{2} \nu^T_{iR} C^{-1} \left( M_R \right)_{ij} \nu_{jR},
\end{eqnarray}
where
$C$ is the charge conjugation matrix. The invariance under $S^I$ implies
\begin{eqnarray}
\left(S^I\right)^T\; M_R \; S^I &=& M_R,
\end{eqnarray}
and thus the symmetric $M_R$ has the form (Eq. \ref{equivalence}):
\begin{eqnarray}
\label{nRm}
M_R &=& \Lambda_R \left ( \matrix{ A_R - B_R   & 0 & 0 \cr
0 & A_R  & B_R  \cr
0 & B_R & A_R  \cr}
\right ).
\end{eqnarray}

Using equations (\ref{seesaw},\ref{nDm},\ref{nRm}), we have the
effective neutrino mass matrix:
\begin{eqnarray}\label{nm} M_{\nu} &=&
-\frac{v^2}{\Lambda_R} \left ( \matrix{ A_{\nu} - B_{\nu}  & 0 & 0 \cr
0 & A_{\nu}  & B_{\nu}  \cr
0  & B_{\nu} & A_{\nu} \cr} \right ),
 \end{eqnarray}
 where
 \begin{eqnarray}
 \label{coef}
 A_{\nu} &=&
 \frac{A_R\,\left(A_D^2 + B_D^2 + C_D^2\right)+ B_R\, \left(C_D^2 - 2\,A_D\, B_D\right)}
 {\left(A_R-B_R\right)\left(A_R+B_R\right)}, \nonumber \\
 B_{\nu} &=& \frac{- B_R\,\left(A_D^2 + B_D^2 + C_D^2\right)- A_R\, \left(C_D^2 - 2\,A_D\, B_D\right)}
 {\left(A_R-B_R\right)\left(A_R+B_R\right)}.
 \end{eqnarray}
Diagonalizing $M_\n$ we get the neutrino mass eigenvalues:
\begin{equation}
 \frac{v^2}{\Lambda_R}\left(A_{\nu}-B_{\nu},\; A_{\nu}-B_{\nu},\; A_{\nu}+B_{\nu} \right).
\label{lnegv}
 \end{equation}

We see here that all possible different patterns of the two degenerate neutrino masses spectrum can be
obtained as follows.
\begin{itemize}
\item{\it Normal hierarchy $\left(m_1=m_2 \ll m_3\right)$ }: It suffices to have
\begin{eqnarray} 0 \leq A_{R,D} \simeq
B_{R,D},\; C_D \ll B_D ,
\end{eqnarray}
for getting a normal hierarchy with
\begin{eqnarray}
A_{\nu} \simeq \frac{A_D^2}{A_R},\; B_{\nu} \simeq \frac{A_D^2}{B_R}.
\end{eqnarray}
We see that one can arrange the
Yukawa couplings to enforce $A_{\nu} \simeq B_{\nu}$, so that to make the smallest neutrino mass $m_1=m_2$ as tiny as one wishes.

\item{\it Inverted hierarchy $\left(m_1=m_2 \gg m_3\right)$}: It is sufficient to have
\begin{eqnarray}
0\leq A_{R,D} \simeq -B_{R,D},\; C_{D} \ll B_D,
\end{eqnarray}
so that one gets an inverted
hierarchy with
\begin{eqnarray}
A_{\nu} \simeq \frac{2 \; A_D^2}{A_R - B_R},\; B_{\nu} \simeq
\frac{-2\; B_D^2}{A_R - B_R}.
\end{eqnarray}
One can arrange the Yukawa couplings to enforce $A_{\nu} \simeq
-B_{\nu}$, so that to make the tiniest neutrino mass $m_3$ small at will.

\item{\it Degenerate case$\left(m_1=m_2 \approx m_3\right)$}: If we have
\begin{eqnarray}
A_{R,D} \gg B_{R,D},\;  B_D \gg C_D,
\end{eqnarray}
then we get
\begin{eqnarray} A_{\nu} \simeq \frac{A_D^2}{A_R},\; B_{\nu} \simeq
\frac{2 A_D B_D}{A_R}.
\end{eqnarray}
which implies $A_{\nu} \gg B_{\nu} $, so that we have a
degenerate spectrum.
\end{itemize}
\vspace{-1mm}
Thus, we see that any pattern occurring in both the Dirac and
the right-handed Majorana mass matrices can reappear in the
effective neutrino mass matrix.

The right handed (RH) neutrino mass term violates lepton number by two units, and the out of equilibrium decay of
the lightest RH neutrino to SM particles can be a natural source of lepton asymmetry \cite{FY}.
This leptogenesis parameter is given by
\begin{eqnarray}
\epsilon \simeq \frac{3}{16\pi
v^2}\frac{1}{(\tilde{M}_\nu^{D\dagger}\tilde{M}_\nu^D)_{11}}\sum_{j=2,3}
\mbox{Im}[\{(\tilde{M}_\nu^{D\dagger}\tilde{M}_\nu^D)_{1j}\}^2]\frac{M_{R1}}{M_{Rj}},
\end{eqnarray}
where $M_{Ri},\; i=1\cdots 3$ are the masses for RH
neutrinos, and $\tilde{M}_\nu^D$ is the Dirac neutrino mass
matrix in the basis where the Majorana RH neutrino mass
matrix is diagonal\footnote{One has to go to the basis where the RH neutrino mass matrix is diagonal because the lepton asymmetry
comes from the decay of the RH neutrino mass eigenstate.}. Explicitly we have
\begin{eqnarray}
(\tilde{M}_\nu^{D\dagger}\tilde{M}_\nu^D)_{11} &=&
\left( 2 \left|C_D\right|^2 + \left|A_D\right|^2 + \left|B_D\right|^2 - A_D\; B^*_D -
A^*_D\; B_D\right),\nonumber \\
(\tilde{M}_\nu^{D\dagger}\tilde{M}_\nu^D)_{12} &=&
\sqrt{2}\;\left( -C^*_D\; A_D + C_D\; A^*_D + C^*_D\; B_D - C_D\; B^*_D\right),\nonumber \\
(\tilde{M}_\nu^{D\dagger}\tilde{M}_\nu^D)_{13} &=& 0,
\end{eqnarray}
 which gives a vanishing  lepton asymmetry . Thus, in this seesaw type mechanism the baryon asymmetry, generated by lepton asymmetry,
 is zero provided $S^I$ is an exact symmetry. Certainly, our symmetry $S^I$ is not exact, and the breaking term (the $C$ part
 in Eq. \ref{tripartite}, which can originate from higher dimensional operators suppressed by a heavy scale) has to be
 added in order to lift the degeneracy among the neutrino masses. In this case one can compute the asymmetry in terms
 of the $S^I$-symmetry breaking parameters. We shall not dwell into this, rather we shall discuss the other phenomenologically motivated possibility
 of leptogenesis in type-II seesaw mechanism.

\section{The neutrino mass matrix and type-II seesaw scenario}
In this scenario we introduce two SM triplet fields $\Sigma_A$, $A=1,2$ which are also assumed to
be singlet under the flavor symmetry $S^I$. The Lagrangian part relevant for the
neutrino mass matrix is
\begin{equation}
{\cal{L} } = \lambda_{\alpha\beta}^{A}\, L_\alpha^T\, C^{-1}\, \Sigma_A\, i\,\tau_2\, L_\beta +
{\cal{L}}(H,\Sigma_A)+h.c.,
\end{equation}
where $A=1,2$ and
\begin{eqnarray}
{\cal{L}}(H,\Sigma_A) &=& \mu_H^2 H^\dagger H + \frac{\lambda_H}{2} {(H^\dagger H)}^2+
M_A\, \mbox{Tr}\left(\Sigma_A^\dagger \Sigma_A \right)+
\frac{\lambda_{\Sigma_A}}{2} \left[\mbox{Tr}\left( \Sigma_A^\dagger \Sigma_A\right)\right]^2
+ \\\nonumber && \lambda_{H\Sigma_A} (H^\dagger H) \mbox{Tr}\left( \Sigma^\dagger_A \Sigma_A\right)
+
{\mu_A H^T \Sigma_A^\dagger i\tau_2 H +h.c.},
\end{eqnarray}
where $H=\pmatrix{\phi^+ \cr \phi^0}$, and
\begin{eqnarray}
\Sigma_A &=& \left ( \matrix{ \frac{\Sigma^+}{\sqrt{2}}  &
\Sigma^0 \cr \Sigma^{++} & -\frac{\Sigma^+}{\sqrt{2}}\cr} \right )_A.
\end{eqnarray}

The neutrino mass matrix due to the exchange of the two triplets, $\Sigma_1$ and $\Sigma_2$, is
\begin{equation} \label{mass}
(M_\nu)^A_{\alpha\beta}\simeq v^2 \left[\lambda^1_{\alpha\beta} \frac{\mu_1}{M^2_{\Sigma_1}} +
\lambda^2_{\alpha\beta}\frac{\mu_2}{M^2_{\Sigma_2}}\right],
\end{equation}
where $M_{\Sigma_i}$ is the mass of the neutral component $\Sigma_i^0$ of the triplet $\Sigma_i ,i=1,2$.

Appropriately, we present some remarks here. First,  the symmetry $S^I$ implies that the symmetric matrices $\lambda_1$ and
$\lambda_2$ have the structure given in eq.~(\ref{equivalence}): $\lam_a = \left ( \matrix{ A_a - B_a   & 0 & 0 \cr
0 & A_a  & B_a  \cr
0 & B_a & A_a  \cr}
\right ),a=1,2$. Second, due to the
`tadpole' term (the $\mu_A$-term) in ${\cal{L}}(H,\Sigma_A)$, which forbids explicitly the
`unwanted' majorons, which would have resultd from the spontaneous breaking of the lepton number, one can arrange the parameters so that
minimizing the potential gives a non-zero vev for the neutral component $\Sigma^0$ of the
triplet. This would generate a mass term for the neutrinos, a procedure which is
equivalent to integrating out the the heavy triplets leading to the same mass formula.
Third, the flavor changing  neutral current due to the triplet is highly suppressed as a result of
the heaviness of the triplet mass scale, or equivalently the smallness of the
neutrino masses.

One can discuss now the baryon asymmetry generated by leptogenesis. We show at present that even though the neutrino Yukawa
couplings are real
it is possible to generate a  baryon to photon density consistent with the observations. In fact, since the triplet
$\Sigma_A$ can decay into lepton pairs $L_\alpha L_\beta$ and $HH$, it implies that these processes violate
total lepton numbers (by two units) and may establish a lepton asymmetry. As the universe cools further,
the sphaleron interaction \cite{KRS} converts this asymmetry into baryon asymmetry. At temperature of the
order $\mbox{max}\{M_1, M_2\}$, the heaviest triplet would decay via  lepton number  violating interactions.
Nonetheless, no asymmetry will be generated from this decay since the rapid  lepton number violating interactions
due to the lightest Higgs triplet will erase any previously generated  lepton asymmetry. Therefore, only when the
temperature becomes just below the mass of the lightest triplet Higgs the asymmetry would be generated.

With just one triplet, the lepton
asymmetry will be generated at the two loop level and it is highly suppressed. We justify this in that one can
always redefine the phase of the Higgs field to make the $\mu$ real resulting in the absorptive part of the self energy diagram
becoming equal to zero. The choice of having more than one Higgs triplet is
necessary to generate the asymmetry \cite{HMS}. In this case,  the CP asymmetry in the decay of the lightest
Higgs triplet (which we choose to be $\Sigma_1$) is generated at one loop level due to the interference
between the tree and the one loop self energy diagram \footnote { There is no one loop vertex correction
because the triplet Higgs is not self conjugate}  and it is given by
\begin{equation}
\epsilon_{CP} \approx -\frac{1}{8\pi^2} \frac{\mbox{Im}\left[\mu_1\mu_2^*
\mbox{Tr}\left(\lambda^1\lambda^{2\dagger}\right)\right]}{M_2^2}
\frac{M_1}{\Gamma_1},
\end{equation}
where $\Gamma_1$ is the decay rate of the lightest Higgs triplet and it is given by
\begin{eqnarray}
\Gamma_1 = \frac{M_1}{8\pi}\left[\mbox{Tr}\left(\lambda^{1\dagger} \lambda^1\right)
+ \frac{\mu_1^2}{M_1^2}   \right].
\end{eqnarray}
If we denote the phases of $A_a-B_a,A_a+B_a,\mu_a$ by $\a_a,\b_a,\phi_a$ ($a=1,2$) respectively, then  by redefining the fields:
 $\Sigma_a \rightarrow e^{-i\a_a}\Sigma_a $, one can remove the phases $\a_a$ in the Yukawa couplings. For  $\mu_a \approx M_{\Sigma_a} \sim 10^{13}\, \mbox{GeV},a=1,2$ (which give a neutrino masses in the sub-eV range) we get:
\begin{equation}
\epsilon_{CP} \approx -\frac{1}{\pi} \frac{2|A_1-B_1||A_2-B_2|\sin(\phi_1 -\phi_2)+|A_1+B_1||A_2+B_2|\sin(\phi_1 -\phi_2+\b_1-\b_2)}
{1+2|A_1-B_1|^2+|A_1+B_1|^2},
\end{equation}
Note that even if $A_a=B_a$, so that to kill the first term in the numerator,  the CP violation responsible of the lepton asymmetry still depends on the relative phases between
$\mu_1, \mu_2$ and/or  $(A+B)_1,(A+B)_2$.

The baryon to photon density is approximately given by
\begin{equation}
\eta_B \equiv \frac{n_B}{s} =\frac{1}{3}\eta_L \simeq \frac{1}{3} \frac{1}{g_*}\kappa \epsilon_{CP},
\end{equation}
where $g_* \sim 100$ is the number of relativistic degrees of freedom at the time when the Higgs
triplet decouples from the thermal bath and $\kappa$ is the efficiency factor  which takes into account the fraction of out-of equilibrium decays and the washout effect. In the case of strong wash out , the efficiency factor can be approximated by ($H$ is the Hubble parameter)
\begin{eqnarray}
\kappa \simeq \frac{H}{\Gamma_1}(T= M_1),
\end{eqnarray}
 With the above numerical values and with an efficiency
 factor of order $10^{-4}$  we get, for $\b_1=\b_2$, a baryon asymmetry:
\begin{equation}
\eta_B\approx  10^{-7}\frac{
\mbox{Tr}\left(\lambda^1\lambda^{2\dagger}\right)}{\mbox{Tr}\left(\lambda^{1\dagger} \lambda^1 \right) + 1}
\sin(\phi_2 -\phi_1).
\end{equation}
Thus one can produce the correct baryon-to -photon ratio of $\eta_B \simeq 10^{-10}$ by choosing
$\lambda$'s of order $0.1$ and not too small relative phase between  $\mu_1$ and $\mu_2$.

\section{{\large \bf Summary and Conclusions}}
We presented here a method to find the most general symmetry implementing the form invariance property satisfied by the neutrino
mass matrix. Applying the method for the tripartite model with two degenerate
masses, we found the underlying symmetry to be the abelian group $U(1)$, which may possibly be enlarged to be $U(1)\times Z_2$, and we have given
a realization of it. The symmetry can be implemented in a complete set up
including charged leptons, and we presented some models to account for the lepton mass hierarchies and the possibility of
 baryogenesis through leptogenesis. The setup can be seen as a first step
approximation, which can be perturbed, with a breaking scale proportional to $C_\nu$ and so to
the neutrino mass splitting (equations \ref{tripartite},\ref{ABC}), so that to lead to
tripartite model without degeneracy.

\section*{{\large \bf Acknowledgements}}
Major part of this work was done within the Associate Scheme of
ICTP. We thank A. Smirnov and S. Petcov for useful discussions. N. C. thanks
CBPF (Brazil), where part of the work has been done, for its hospitality and acknowledges support from TWAS.

\end{document}